# A new microscopy for imaging retinal cells

Timothé Laforest[1*a], Dino Carpentras[1*], Mathieu Kunzi[1], Laura Kowalczuk[2,3], Francine Behar-Cohen[2,4], Christophe Moser[1]

1 School of Engineering, LAPD, École Polytechnique Fédérale de Lausanne (EPFL), Lausanne, Switzerland

2 Faculty of Biology and Medicine, University of Lausanne, Lausanne, Switzerland

3 Jules-Gonin eye hospital, Fondation Asile des aveugles, Lausanne, Switzerland

4 Inserm, U1138, Team 17, From physiopathology of ocular diseases to clinical development, Paris Descartes University, Sorbonne Paris Cité, Centre de Recherche des Cordeliers, Paris, France

*authors contributed equally to this work
a: timothe.laforest@epfl.ch

**Abstract**

The evaluation and monitoring of cells health in the human retina is crucial and follow time course of retinal diseases, detect lesions before irreversible visual loss and to evaluate treatment effects. Towards this goal, a major challenge is to image and quantify retinal cells in human eyes in a non-invasive manner. Despite the phenomenal advances in Optical Coherence Tomography (OCT) and Adaptive Optics systems, *in vivo* imaging of many of these cells is limited by the fact that cell contrast in reflection is extremely low. Here, we report on a major advance by proposing and demonstrating a radically different method compared with OCT to visualize retinal cells with high contrast, resolution and an acquisition time suitable for clinical use. The method uses a transscleral illumination which provides a high numerical aperture in a dark field configuration. The light backscattered by the Retinal Pigment Epithelium (RPE) and the choroid provides a forward illumination for the upper layer of the retina, thus providing a transmission illumination condition. By collecting the scattered light through the pupil, the partially coherent illumination produces dark field images, which are combined to reconstruct a quasi-quantitative phase image with twice the numerical aperture given by the eye's pupil. The retrieved quasi-quantitative phase images (QPI) of cells in *ex vivo* human and pig's eyes are validated with those taken with a standard QPI system. We then apply the technique *in vivo* on human eye, without pupil dilation, we demonstrated the feasibility to image retinal cells up to the retinal pigment epithelium within a few seconds.

## Introduction

Retinal diseases are the major cause of blindness in industrialized countries and while tremendous effort is made to develop novel therapeutic strategies to rescue retinal neurons, optimal means to evaluate the effects of such treatments is still missing.

In order to understand the difficulties in retinal imaging by optical means, it is appropriate to start with a description of the retinal structure. The human retina is composed of several distinct layers of transparent neuronal, vascular and glial cells, with a thickness ranging from 150 to 300 um depending on the area of the retina, and on the retinal pathology of the subject [1]. These higly transparent layers allow the photons entering the eye to go through them before reaching the photoreceptor cells where phototransduction takes place. As a consequence, the retina structure has very low light reflectivity. Hence, every optical method to image the retina is constrained by this low light reflectivity.

Optical Coherence Tomography/Microscopy (OCT/OCM) is a well-established depth resolved imaging method in ophthalmology. The method is based on interfering a reference beam with the beam reflected by the weak variation of refractive index that occurs in the stratified structure of the retina [2, 3]. OCT is thus a technique used to monitor retinal structure alterations during clinical follow up, for example the central macular thickness. Minute changes in cell morphology which maybe present during the early stages of disease progression cannot (yet) be quantified by OCT. Various diseases of the eye are associated with characteristic morphological changes in different layers of the retina. Analysis of individual layers of the retina is therefore becoming increasingly important with the advent of new therapeutic approaches. Thus, the development of methods that can extract and quantify such small morphological changes in retinal cells at different depth is therefore of considerable interest.

Other optical imaging techniques routinely used in ophthalmology research are wide field and linear confocal imaging coupled with adaptive optics. These techniques are employed to resolve cone cells in the fovea. This is possible because the photodetector cell assembly is highly reflective (~2%) [4,35]. Despite these tremendous developments in optical imaging techniques, several structures of the retina can still not be imaged *in vivo* [25]. On the other hand, phase imaging techniques can image transparent cells with high contrast. However, phase imaging techniques such as differential interference contrast [19-20] and digital holographic microscopy (DHM) [21] are limited to a transmission arrangement, preventing its applicability to *in vivo* retinal imaging.

Recent works proposed optical methods to observe transparent cells and microvasculature with phase contrast in animal retinas. Toco et al [7-8] used confocal dark field imaging of microvasculature close to the optic disc thanks to an offset confocal aperture through the pupil. Three other studies reported results based on split-detector, combined with an adaptive optics scanning laser ophthalmoscope (AOSLO) [9-11]. The split-detector allows obtaining differential measurements, thus removing background and absorption terms in the image. These approaches produced phase contrast images of ganglion cells and microvasculature in the mouse retina. Recently Rossi et al [32] showed, with the offset aperture method, high contrast images of ganglion cells in monkeys. In this latter study, however, because of the limitation in retinal illumination level due to safety concerns in humans, the obtained image contrast of ganglion cells is poor which would make their use difficult for chronic evaluation in a clinical setting. The contrast of phase objects is ultimately limited by the transpupillary illumination which produces specular light and by the relatively low frame rate of the scanning system. In another study Liu et al. [36] used an adaptive optics system combined with OCT to obtain high resolution imaging of the neuroretina. Such a system provided enough resolution to observe single neurons and enough contrast to detect signal from the organelles of the cells. Organelles move within the soma, thus, by averaging more than 100 images over a time span of approximately 10 minutes, they showed that it was possible to obtain images of ganglion somas. Despite the great quality of the result, this imaging technique is limited by organelle motility, making the acquisition process too slow for clinical use.

Here, we present a method that circumvents the time limitation due to organelle's motility and the contrast limitation of off-axis detection methods, to provide high-resolution phase imaging of the retina. Moreover, we demonstrate that the method produces a quasi-quantitative measure of phase, which has been shown to be powerful information to evaluate cell health in addition to morphology [21].

Our approach is based on transscleral illumination with multiple oblique directions to record dark field images of the retina, and then combine them to reconstruct a quantitative phase image. Transscleral illumination has been used for decades for diagnostic in oncology and in retinal surgery of the eye but never for cell imaging. By passing through the sclera, light produces a high angle oblique illumination of the retina which provides a larger illumination numerical aperture than the collecting aperture of the pupil [16]. This light is back-scattered by the Retinal Pigment Epithelium (RPE, the deepest layer of the retina) and the choroid which produces a secondary illumination propagating through the transparent retinal layers (see fig 1). This illumination condition by backscattering from tissue is akin to the endoscopic probe of Mertz et al [13,14]. The processing of two asymmetric dark field images removes the absorption and background terms of the collected beam by computing the differential phase contrast image. Knowing the illumination transfer function, and assuming a weak phase object, it is possible to recover the quantitative phase information of the sample [15-18, 30-31]. If the shape of the illumination function is known, it is still possible to reconstruct a phase profile that is proportional to the original one. Such an image is called quasi-quantitative in this work. Since backscattered light is used as an effective transmission illumination, an accurate modeling of the illumination diffusion is also required. To reach *in vivo* cellular resolution in the retina, one must undo the optical aberrations of the eye. Several aberration correction methods exist. A first method relies on using adaptive optics, which is the method we used here in this work [22]. A second option is to correct the aberration computationally using a guide star algorithm [23] or a sharpness based criteria algorithm [24].

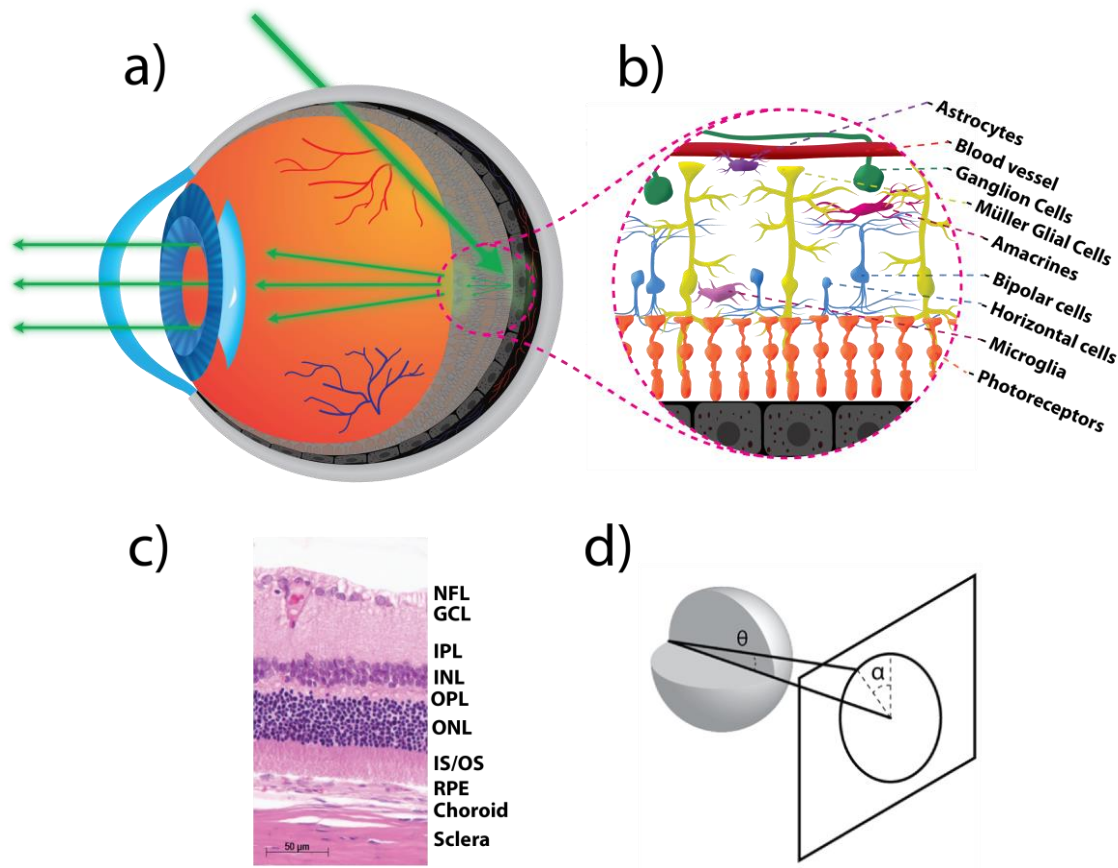

**Figure 1. Illumination of the retinal layers provided by transscleral illumination.**

(a) Light transmission into the eye. The light is first transmitted through the sclera, the retinal pigment epithelium (RPE) and the neuroretina. After travelling through the vitreous humor, it impinges on the RPE layer. Here backscattering off, the RPE generates a new illumination beam. This secondary illumination provides a transmission light propagating through the transparent medias: the neuroretina and its translucent cells, the vitreous, the eye lens, and then, the anterior segment of the eye. (b) Illustration of the translucent vascular, neuronal and glial cells of the retina. (c) Histological cut of the retinal layers, the choroid and the sclera. (d) Azimuthal angle θ and polar angle α.

The paper is organized as follows: in section 1, we introduce the method and the system. In the "ex vivo demonstration section", we first assess the quantitative phase measurement of the proposed technique in an *ex vivo* apparatus mimicking the apparatus used for *in vivo* imaging. We verified the doubling of 2D spatial resolution compared to coherent illumination using the USAF intensity test target. Next, the quasi-quantitative phase images retrieved with the *ex vivo* set-up are compared with those obtained from a digital holographic microscope

(DHM). We also compare our phase images with reflectance confocal microscopic images to demonstrate the improvement of contrast for transparent cells. We then present experimental 3D depth resolved results of the proposed phase imaging technique using *ex-vivo* samples of human and pig eyes. In order to unequivocally associate the phase images with known cells, we compare phase images with fluorescent images. Finally, in the "in vivo demonstration section", we apply the technique to living human eyes, obtaining comparable results with the *ex-vivo* study without use of artificial pupil dilation.

**Detailed description of the method**

Transscleral illumination of the fundus has a higher numerical aperture than what is obtainable via illumination through the pupil. In Fourier domain, this is equivalent to a shift of the illumination towards higher spatial frequencies, meaning exciting higher spatial frequencies. In addition, oblique illumination coupled with an imaging system which captures images through the pupil, but does not collect the superficial specular reflection, produces dark field images of the fundus, thus providing a high Signal to Noise Ratio (SNR) allowing detection of the dim light reflected from the fundus deep layers. The forward scattered incoherent light by the sclera illuminates the fundus uniformly. Additionally, the use of partially coherent light gives a system's bandwidth that is twice the bandwidth obtainable with coherent light. A dark field image can be obtained with just one illumination source. Phase imaging is performed thanks to the following procedure: at least two pictures, I(α) and I(α+180) captured with two different asymmetric illumination polar angles α and α+180° are required. One can then obtain the differential phase contrast (DPC) image according to:

$$I_{DPC} = \frac{I(\alpha) - I(\alpha + 180°)}{I(\alpha) + I(\alpha + 180°)}$$

By combining different DPC images of the same object, it is possible to obtain a quantitative phase image of that object [15]. The illumination direction is an important parameter in the phase reconstruction process.

**Ex-vivo demonstration**

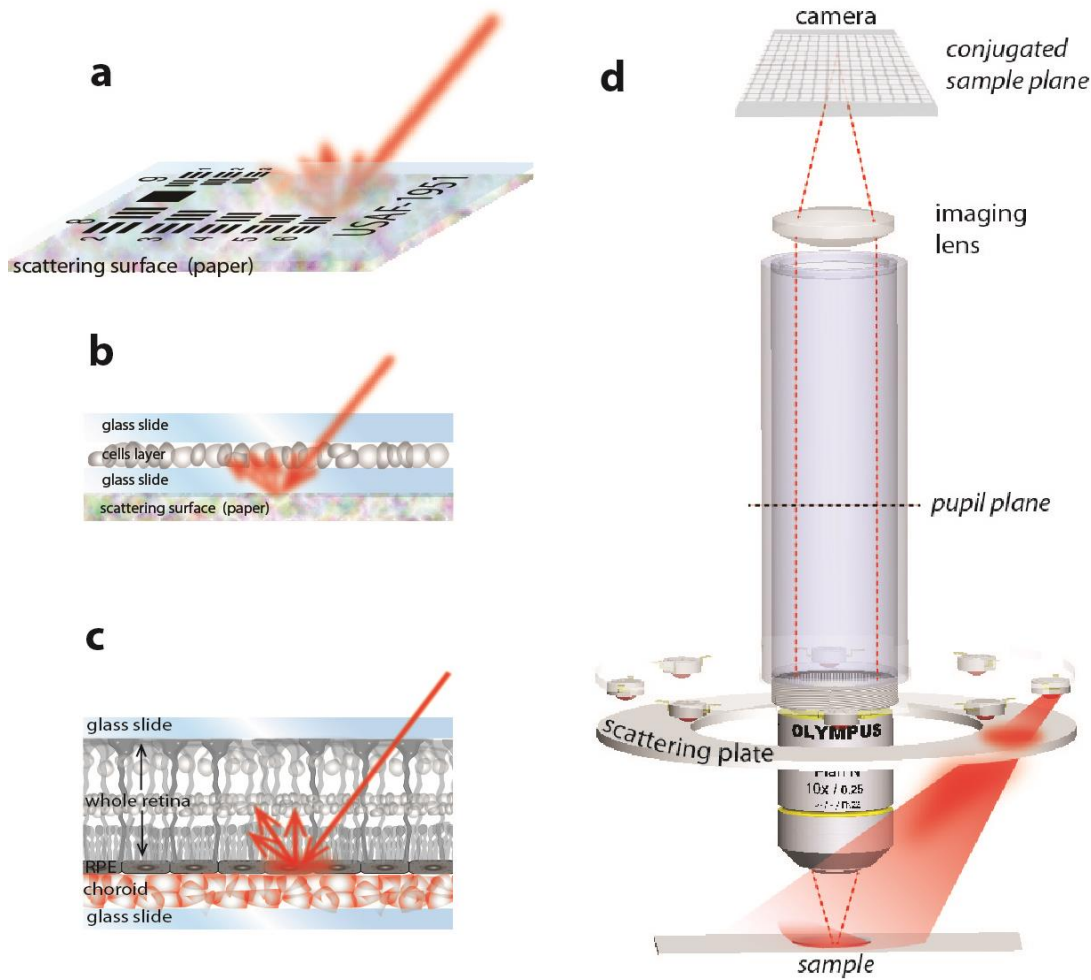

**Figure 2. Experimental set up for *ex vivo* observations. (a-c) Schemas of the imaged samples. (a) Intensity USAF** target sample mounted on top of a scattering surface (paper) for assessing the modulation transfer function of the optical system. Incoming light back-scatters from the paper. **(b) Frontal frozen sections of a pig neuroretina**, mounted on a glass slide. A scattering surface (paper) is attached to the bottom side of the glass slide in order to provide back scattering light. **(c) Flat-mounted "retina - choroid" complex on a glass slide.** The light back-scatters from the RPE-choroid layer **(d) Experimental microscope for phase imaging**. LED light illuminates the sample with multiple directions. The backscattered light effectively becomes a forward beam that illuminates the transparent phase layers in transmission geometry before reaching the microscope objective (10x, 0.25NA). An imaging lens makes an image of the sample on a camera.

The experimental setup is illustrated in Fig 2.d. A 0.25 NA objective is selected to match the maximum numerical aperture of a fully dilated human eye (0.24).

The Light Emitting Diodes (LEDs) have a center wavelength of 650 nm and 50 nm bandwidth. To mimic the scattering effect of the sclera, a paper sheet is placed between the LEDs and the samples as illustrated on Fig 2.d. We first measured the modulation transfer function (MTF) of our system using the groups 8 and 9 of a USAF intensity target. Fig. 3 a shows that the measurement fits well with the ideal MTF of an incoherent illumination having a cutoff frequency of 2NA/lambda. The graph shows that the 780 nm wide bars (i.e. 1.56 µm period) are well resolved with a 0.25 NA objective and 650 nm wavelength light, thus confirming the doubling of the resolution compared to a coherent imaging system.

Then, we performed imaging of frontal frozen sections of pig neuroretina. A scattering surface (paper) has been attached to the bottom side of the glass slide in order to simulate the back scattered-light, as shown in Fig 2.b. These sections were imaged with both a digital holographic microscope (DHM) and the proposed phase imaging method. DHM produces true metrological phase images, allowing a quantitative comparison with the proposed method. The images from the DHM are taken with a 0.4 NA microscope objective and with coherent light illumination. Since our method uses incoherent light, the cut-off spatial frequency is equal to $2NA/\lambda = 0.5/\lambda\,[m^{-1}]$. Thus, our phase imaging method produces images with similar resolution as the 0.4 NA DHM images taken with coherent illumination and with a sharp cut-off spatial frequency of $NA/\lambda = 0.4/\lambda\,[m^{-1}]$.

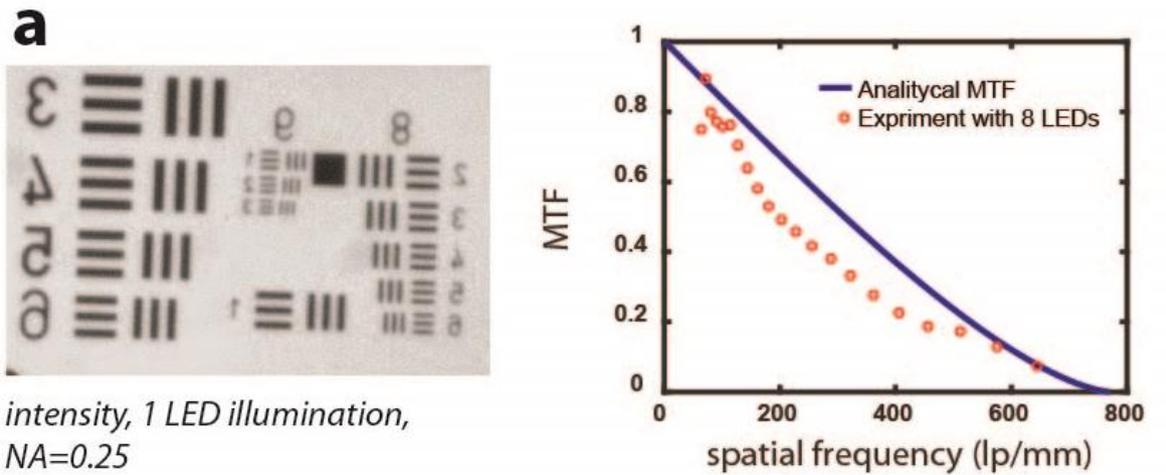

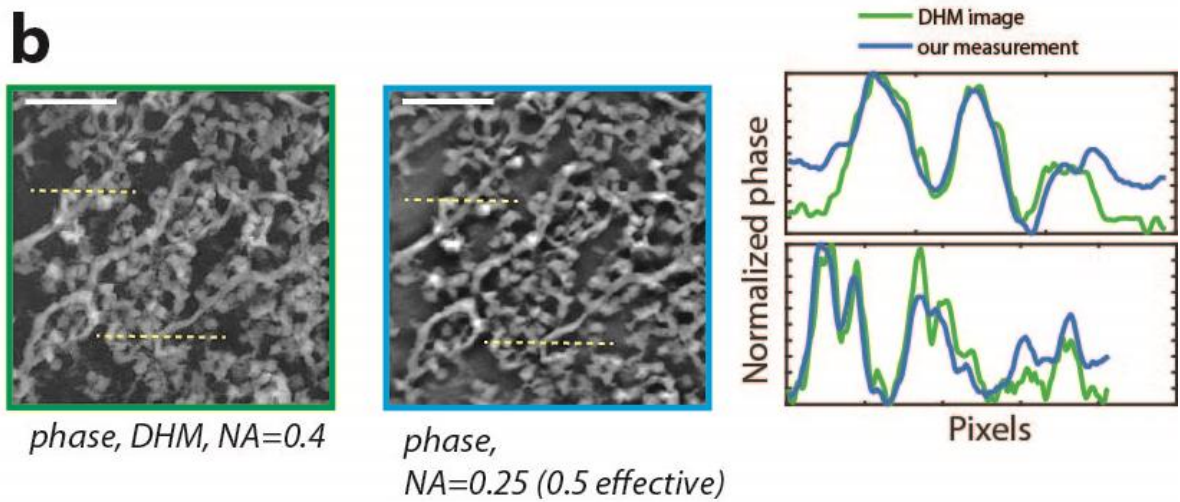

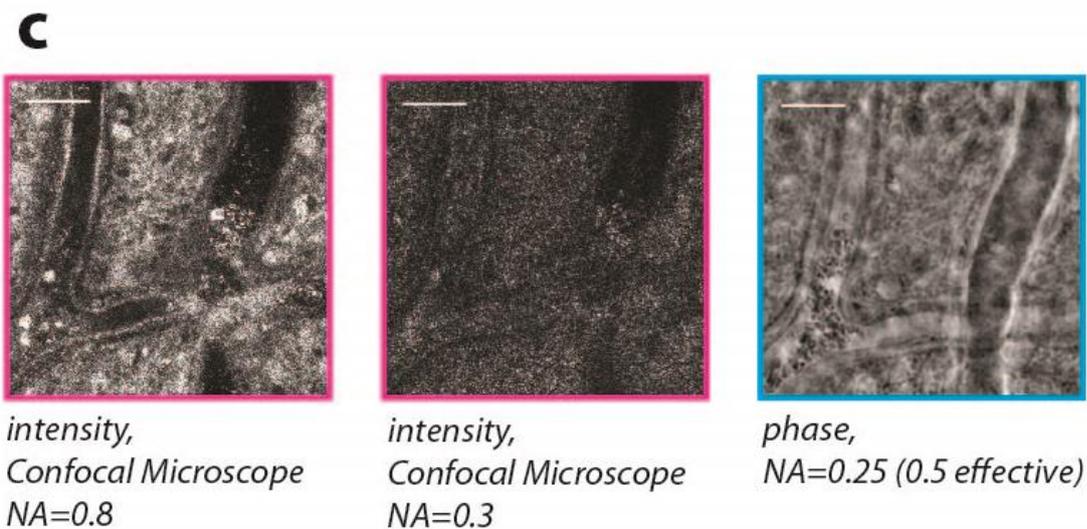

**Figure 3. *Ex vivo* imaging of the USAF intensity target (a), of a pig neuroretina section (b) and of a human flat-mounted retina-choroid complex (c). (a) Phase imaging of the USAF intensity target.** (left) Dark field image of the USAF intensity target obtained with a single LED illumination. (Right) Plot of the theoretical and experimental modulation transfer function from the image. Red dots correspond to the contrasts of elements of groups 8 and 9 of the USAF. **(b) Imaging of a 10 um-thick frontal frozen section of pig neuroretina at the inner nuclear layer level.** (Left) Phase image obtained with a digital holographic microscope. (Center) Phase image obtained with the proposed method. (Right) Horizontal cross section plot of quantitative phase comparison. Scale bars = 30 um. **(c) Imaging of a human flat-mounted fixed retina-choroid complex at the nerve fiber layer level.** (Left-middle) Reflectance confocal images taken with 0.8 NA and 0.3 NA objectives respectively. (Right) Corresponding phase image with a 0.25 NA objective, corresponding to a 0.5 effective NA. Scale bars = 50 um.

Furthermore, we compared our phase imaging modality to reflectance confocal microscopy, using a human retina-choroid complex. This sample was prepared as shown in Fig. 2.c, with the back illumination provided by the scattering of the RPE-choroid layers. In order to compare the images, the region was selected using blood vessels as land marks. Fig 3c shows the intensity images taken with 0.8 NA (left) and 0.3 NA (center) objectives of the confocal microscope. The phase image (right) recorded at the same depth exhibits better SNR and resolution compared to the 0.3 NA confocal microscope image. In addition, compared to the 0.8 NA image, the phase image makes other features visible, as blood cells into the vessel and other retinal cellular structures out of the vessel.

Finally, we demonstrated depth resolved phase images on a pig retina - choroid complex (see Fig 2.c). Here, the back illumination was also provided by the scattering of the RPE-choroid layers. Depth scanning was obtained by moving the sample axially, following z-axis. The 0.25 NA objective used in the experimental setup gives a depth of field of 8.5 µm, which allows imaging every layer of the 100 µm-thick retina. The imaged region on the retina was the *area centralis* [24]. Phase imaging allowed the visualization of all retinal layers which were digitally tagged to highlight the cells (Fig. 4).

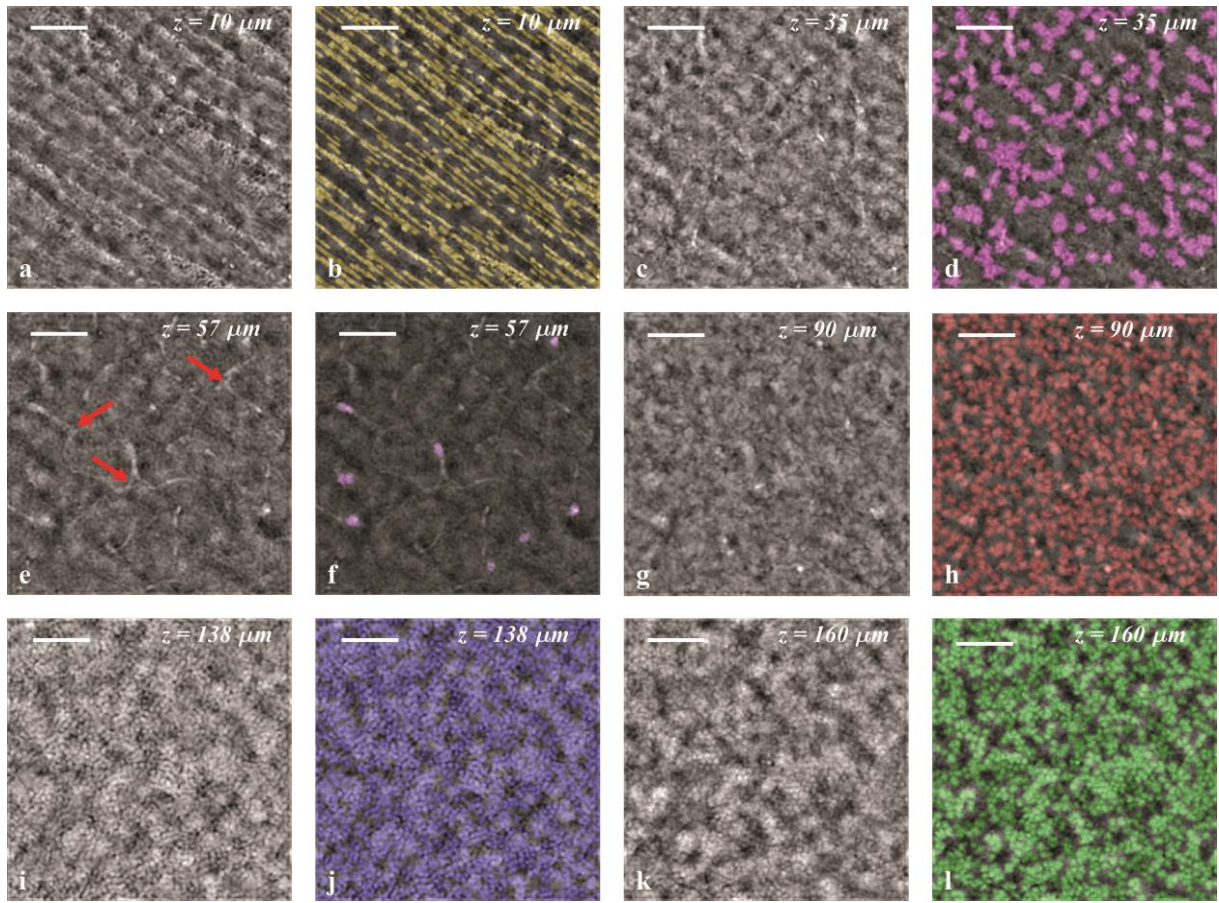

**Figure 4. Scan in depth of a flat-mounted fixed pig retina-choroid complex using phase imaging**. (a, c, e, g, i, k) raw phase images. (b, d, f, h, j, l) phase images with digitally tagged cells and structures. (a,b) Nerve fiber layer, z = 10 um. (c,d) Ganglion cells layer, z = 35 um. (e, f) Inner plexiform layer, with microglia (red arrows); z = 57 um. (g,h) Inner nuclear layer, z = 90 um. (i, j) Outer nuclear layer, z = 138 um. (k, l) Inner / outer photoreceptors segments interface, z = 160 um. Scale bars = 50 um.

The raw images of each distinct layer have been processed in order to tag cell nuclei, and thus to quantify cell densities. The density analysis for the GCL, INL, ONL and PR layer is reported on figure 5 with blue curves. The data from the literature of pig retinal cell densities were compared to our experimental density values at the GCL [26] and the photoreceptors layer [27, 28] levels. We measured a ganglion cell density of 2'260 cells/mm$^2$ and a mean cone densities of 20'930 cells/mm$^2$ which are consistent with the ganglion cell densities (1500 to 4000 cells/mm$^2$, [26]) and the cones densities (19'000 [27] to 22'600[28] cells/mm$^2$) reported in the literature.

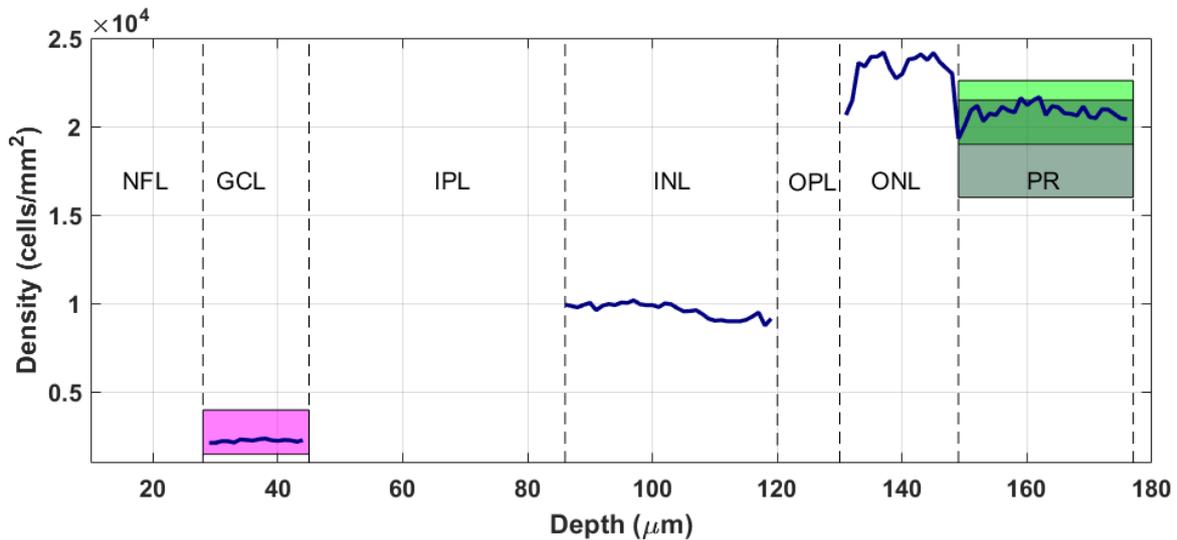

**Figure 5. Cell densities over depth. Blue curves:** measured data in this study. **Magenta**: density range of ganglion cells from [26]. **Green**: density range of cones from [27]. **Dark green**: density range of cones from [28]. GCL, ganglion cells layer; INL, inner nuclear layer; IPL, inner plexiform layer; NFL, nerve fiber layer; ONL, outer nuclear layer; OPL, outer plexiform layer; PR, interface between the inner and outer segments of the photoreceptors.

In order to identify the retinal cells that our method is able to image, normal rat neuroretinas were immunostained to specifically detect pericytes in fluorescence. By imaging the same region of one neuroretina with fluorescent confocal microscopy and with our experimental phase microscope, we correlated both imaging modalities. The fluorescent images were acquired with a 0.8 NA objective, and the phase images were generated using a 0.4 NA objective (yielding a comparable resolution with effective NA of 0.8). The correlation image analysis demonstrates that the phase imaging method allows the detection of pericytes around retinal capillaries (Fig. 6).

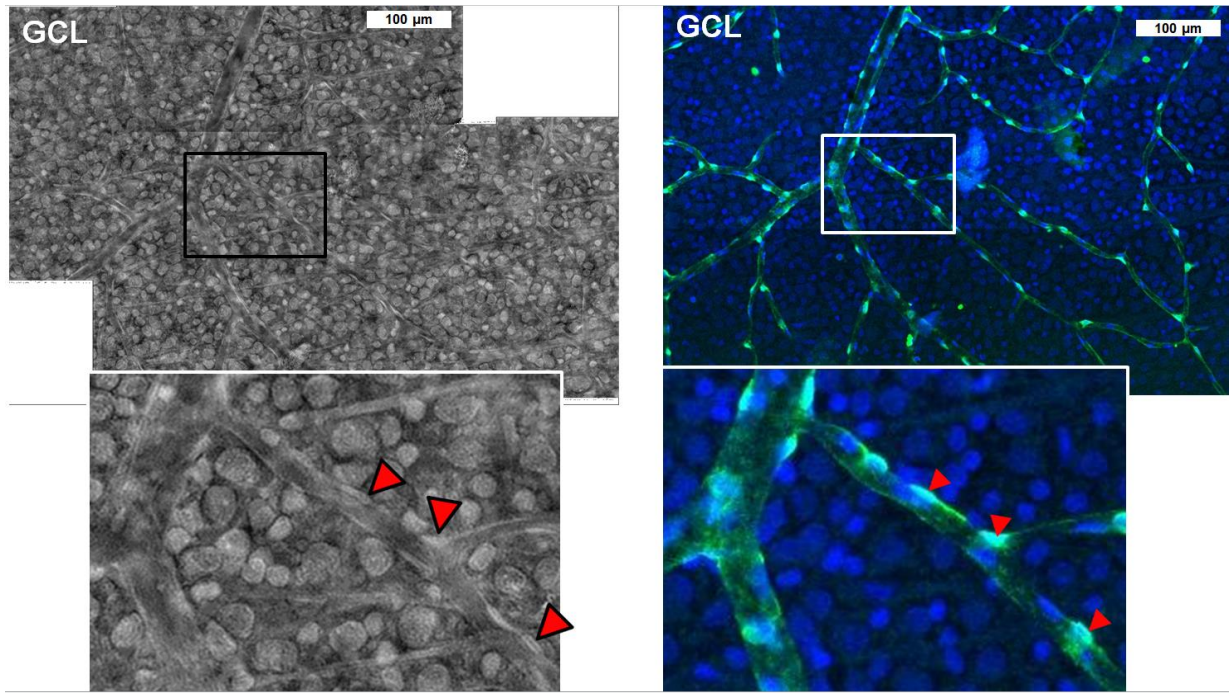

**Figure 6. *Ex vivo* observation of a rat neuroretina at the ganglion cell layer (GCL) level**. Retinal phase images (left) were compared with fluorescence microscopy (right, same location). Cells highlighted with red arrowheads in the phase image are exactly localized where the pericytes (NG2-green) are detected in the fluorescence image. Nuclei are stained in blue with DAPI.

### *In-vivo* demonstration

For the *in vivo* demonstration, the angled illumination on the retina is produced by two laser diode beams illuminating the sclera successively on the temporal side of the iris and the nasal side of the iris, at a 15 mm distance from the pupil center (see Fig. 7). The obtained angle of illumination is therefore approximately 35°. The illumination arm is also composed of a yellow fixation target, of a 780 nm Superluminescent Diode (SLD) providing a point source at the retina plane and of an infrared LED for illuminating the pupil uniformly. The imaging arm integrates a pupil camera, an adaptive optics feedback loop (Mirao 52e deformable mirror and associated wavefront sensor, Imagine Eyes). The retinal sCMOS camera (ORCA Flash, Hamamatsu) collects the light from the transscleral illumination to image the eye fundus with a sampling of 1 µm/pixel.

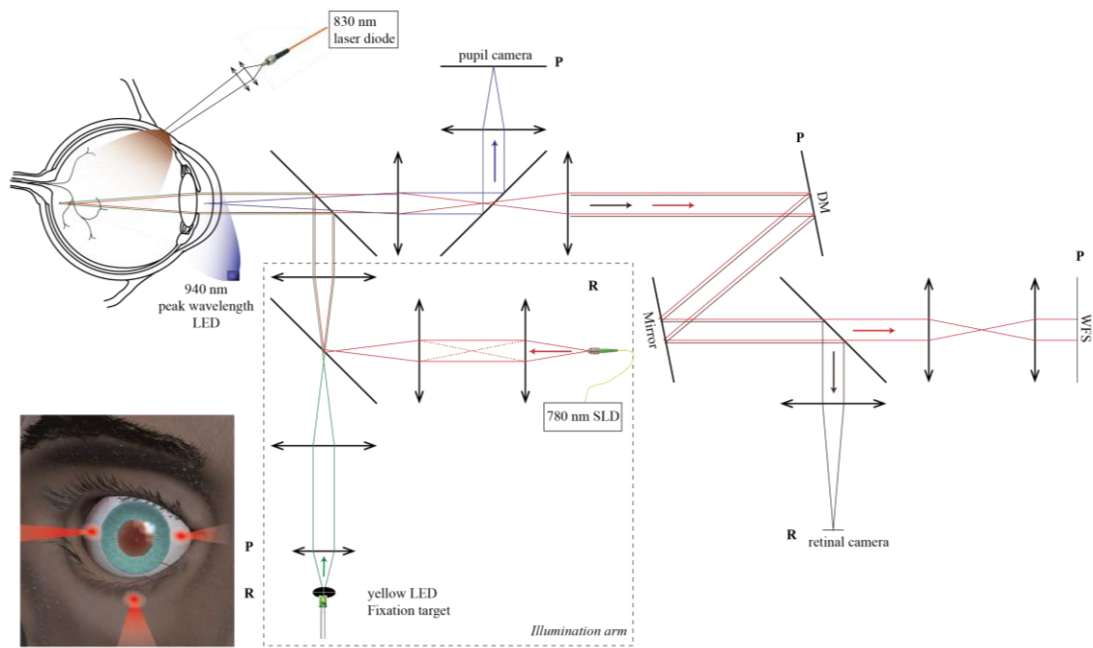

**Figure 7. Schematic of the *in vivo* phase imaging system.** The digital photography represents transscleral and transpalpebral illuminations by means of a focused beam on the sclera or on the skin of the inferior eye lid. Light is then transmitted inside the eyeball. After scattering off the eye fundus, the light passing through the retinal cell layers is collected through the pupil of the eye. The optical setup includes an adaptive optics loop to correct the aberration of the eye. DM, Deformable mirror; LED, light-emitting diode; SLD, superluminescent diode; WFS, wavefront sensor

Once the position of the eye is aligned on the setup, the retinal camera, mounted on a translation stage, is adjusted to focus on a retina layer. The depth sectioning is here limited by the depth of field, which is given by the pupil size and the aberrations of the eye. For instance, for an aberration free 6 mm pupil, the depth of field is 27 µm, for a lateral resolution of 2.9 µm.

A sequence of approximately five thousand images, covering twelve different layer depth in the retina, is acquired by turning successively the laser diode spots ON and OFF with an 8 ms exposure time and a frame rate of 16 Hz. The distance between successive depths is 12 µm (half of the depth of field) to have a best focused image of each layer. A single layer takes 25 seconds to image which is suitable for clinical application. The images from each illumination point are averaged after computing the shift due to eye motion. The phase image is reconstructed thanks to the regularization process described in [16], using the proper illumination function of the back scattered light.

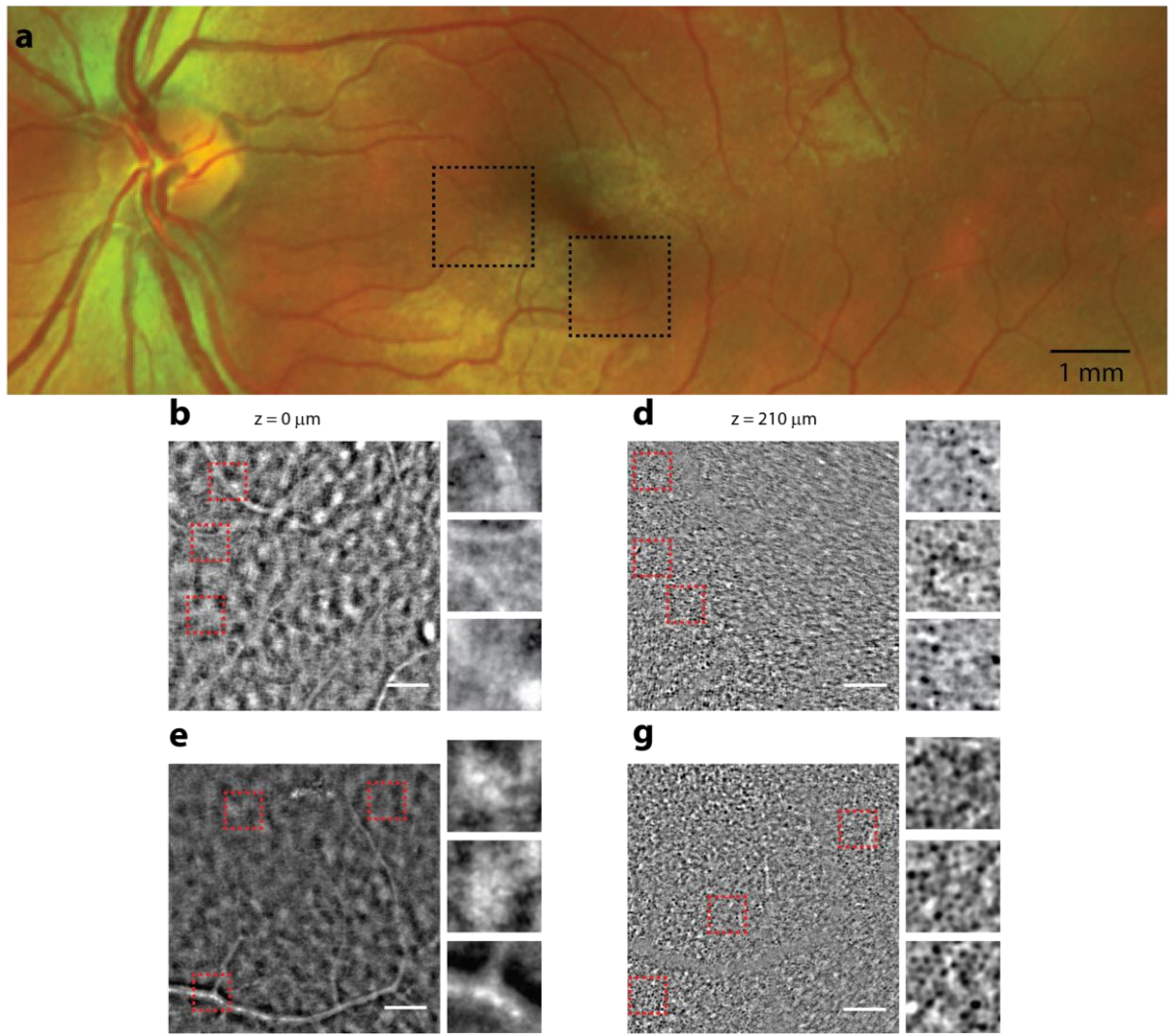

**Figure 8.** *in vivo* retina imaging of one healthy volunteer. (a) Left eye fundus on scanning laser ophthalmoscope (Optos). (b-g) *in vivo* phase images of the two areas highlighted with black squares on the eye fundus, at 1.5 mm eccentricity from the fovea, at three different depths into the retina: nerve fiber layer (b, e) and retinal pigment epithelium (d, g).5 mm pupil. Scale bars = 200 μm. The patches are 125 μm x 125 μm.

*In-vivo* phase images were acquired on a healthy volunteer (see supplementary materials for ethical approval details) with a non-mydriatic pupil of 4 to 6 mm diameter, in dark room conditions.

Fig 8a shows the eye fundus of the which was then observed with our experimental *in vivo* phase imaging device (Fig. 8 b-g). The processed images from one illumination point show the nerve fiber layer (Fig. 8b, e), and the RPE layer(Fig. 8d, g) acquired with a shift of 210 µm deeper than the NFL, . Thanks to oblique illumination, the RPE layer appears well contrasted avoiding the Stiles-Crawford effect [38] that occurs in standard transpupillary illumination [37]. We attribute the high RPE contrast to the oblique illumination which is highly transmitted through the photodetector layer and, at the same time, the boundaries of the RPE cells diffuse light which is collected by the pupil. Fig. 9 shows the *in-vivo* RPE cells at different excentricity from the fovea. The graphs on the right in Fig 9. show the RPE density density and area. These values are comparable to previously reported values [6], as well as the row to row spacing extracted from the Fourier analysis, shown in Fig. 10. Fig 11 shows the enhancement of the contrast over noise ration (CNR) with respect to the number of frame averaged. Fig 12 is a comparison of dark field images of RPE cells obtained *in-vivo* on human and *ex-vivo* on pig.

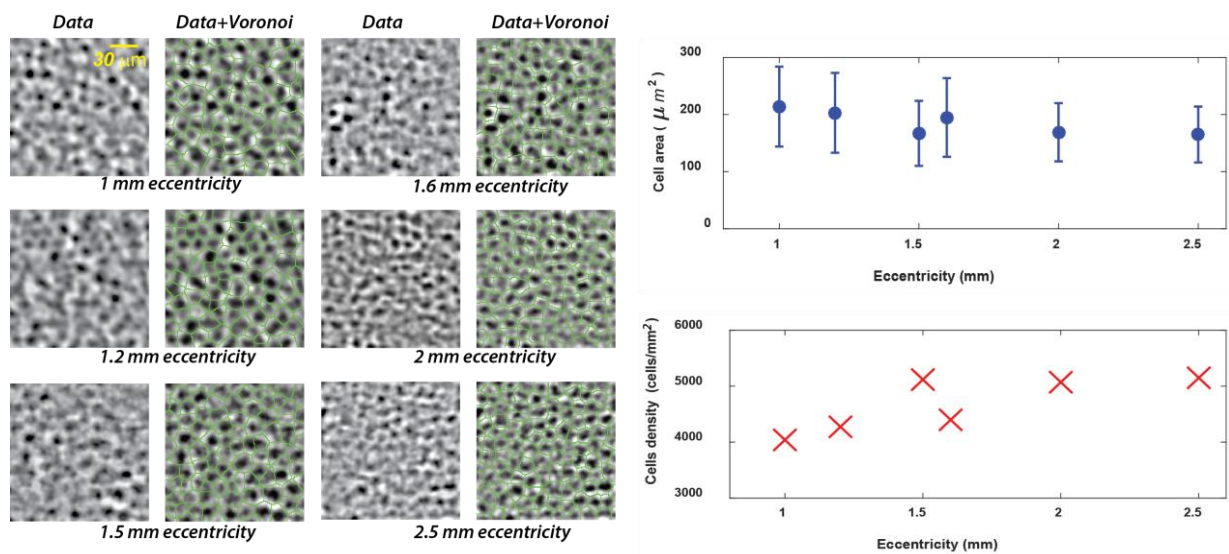

**Figure 9.** Density and area analysis of the RPE cells. Data and overlap with a Voronoi filter at different eccentricities (Left). Cells density and Area (Right).

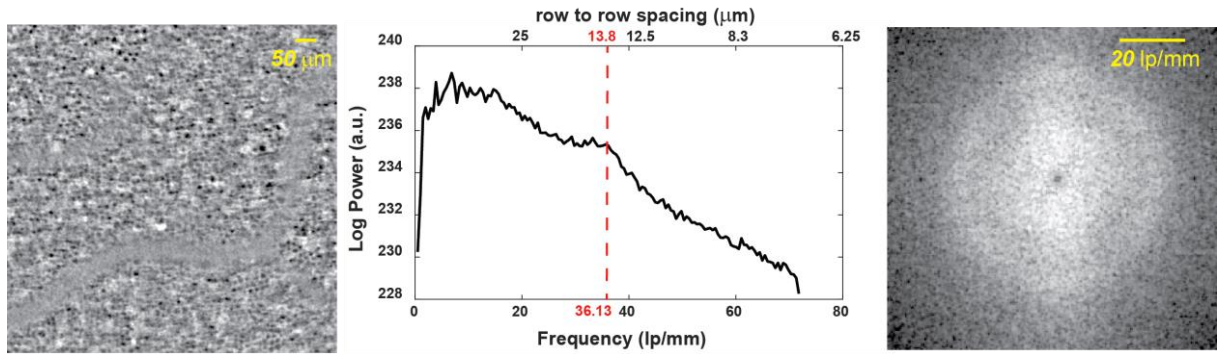

**Figure 10.** Fourier analysis of the RPE layer image centered at 7° from the fovea. Dark field image high pass filtered to enhance the visualization (Left). Axial profile of the spectrum (Center). Spectrum of the left image cropped at 6.8 µm row to row spacing (Right). The peak signature of the RPE signal is located at 13.8 µm, which is consistent with the values found in the literature.

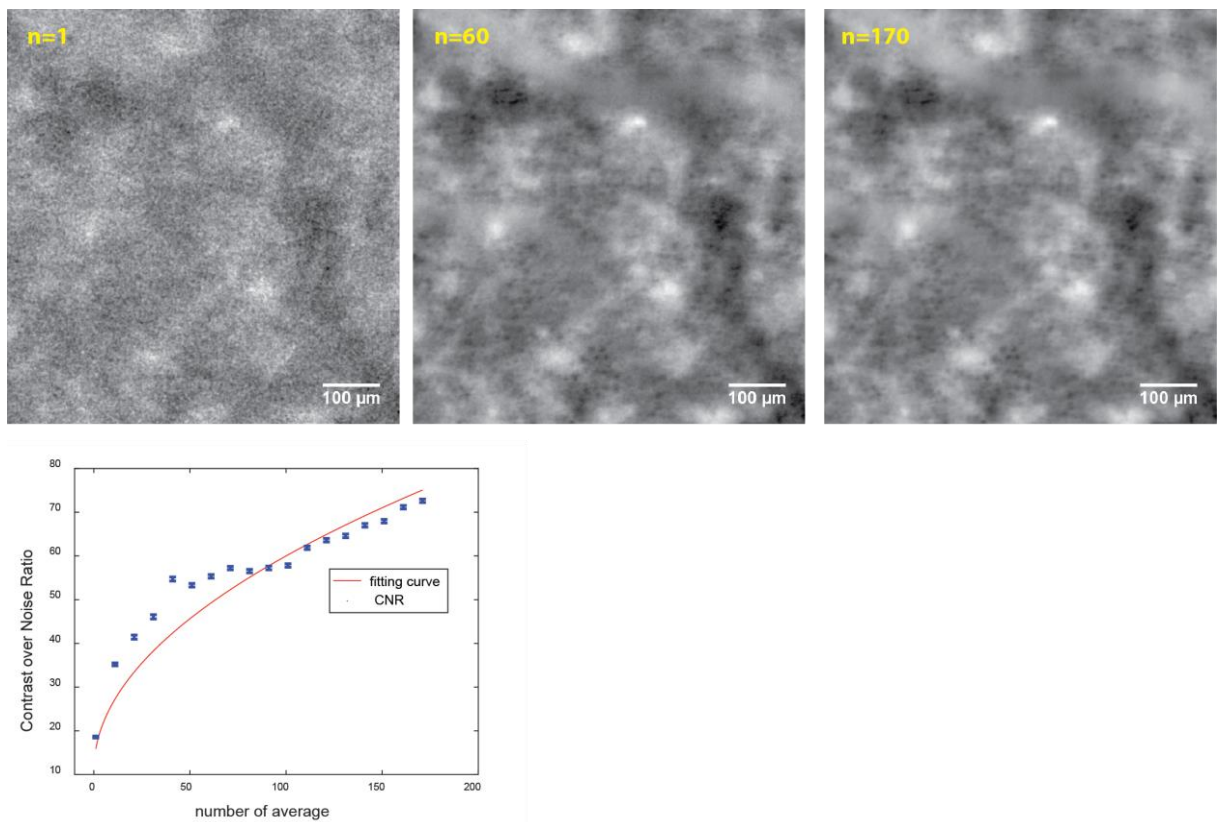

**Figure 11.** Averaging of the RPE layer image centered at 7° from the fovea. Dark field images are aligned before averaging.

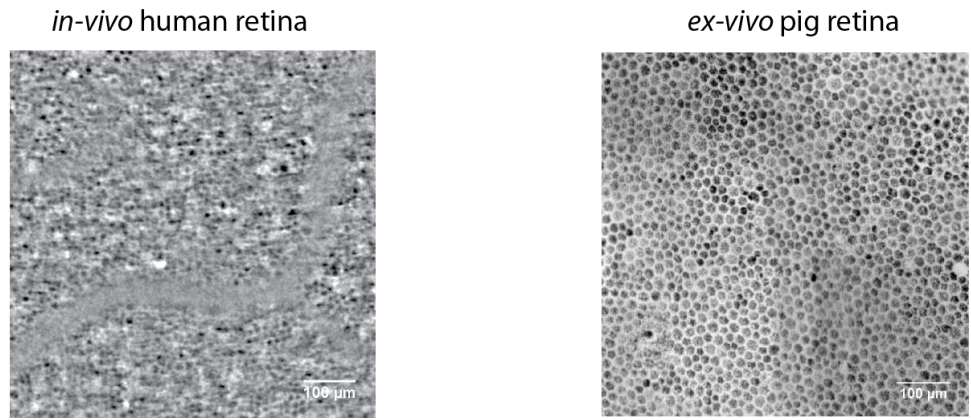

**Figure 12. *In-vivo* and *ex-vivo* comparison of dark field images of the RPE layer**. *In-vivo* human RPE with ~5 mm diameter pupil, illumination with 830 nm peak wavelength laser diode (Left). *Ex-vivo* pig RPE registered with 0.4 NA 20X microscope objective illumination with 850 nm peak wavelength LED (Right).

**Discussion**

The present study reports a new microscopic concept and instrument for *in vivo* phase imaging of all the retinal layers with high contrast and cellular resolution. We first validated our phase imaging method *ex vivo* through the observations of human, rat and pig retinas, performing cell density measurements and correlative image analysis with digital holography microscopy and with confocal microscopy. Finally, during *in vivo* examination of one healthy volunteer using only two transscleral illumination beams, we demonstrated the feasibility to image the retinal cells within seconds, from inner layer up to the deeper RPE cells layer without pupil dilation.

 This novel method takes advantage of the relative transparency of the sclera and choroid/ RPE complex, which allows 30% of light transmission to the retina. Since the angle of illumination is larger than the maximum angle of the offset aperture, it provides higher contrast of the phase objects into the retina, subsequently improving the contrast of the dark field image. Moreover, the use of oblique flood illumination provides a dark field condition, which enables fast single shot full-field image acquisition that does not depend on biological processes [6,36].

Importantly, this illumination system presents the advantage to reduce the light power into the eye as compared to existing technology. Indeed, assuming that 30% of the incoming beam on the sclera is transmitted inside the eye and the transscleral light scatters only on the posterior globe of the eye, an energy of 36 µJ/cm2 is required for taking a picture at a given depth. This energy is comparable to the one required for AO flood transpupillary illumination [39]. In the AOSLO set up presented by Rossi et al for retinal cell imaging in human, the light power at the cornea was approximately 290 µW. [32] The light is then focused on the retina to obtain one pixel on the reconstructed images. For their *in vivo* human measurements, 6 to 8 frames were averaged. For AO-OCT device allowing imaging the ganglions [36], Liu et al reported an incident power at the cornea of 430 µW with A scan and B scan rates of 500 KHz and 1.1 KHz and a sampling of 1 µm/pixel in both lateral directions. This leads to an energy of 86

mJ/cm2 for one *en face* view. As a comparison, commercial clinical OCT (10 µm/pixel sampling) instruments use an energy of approximately 10 mJ/cm$^2$ for one *en face* view. In our study, we took advantage of recording single-shot 4 megapixels image. Hundred images were averaged to obtain a dark field image. Contrast of the transparent cells is increased because of the dark field configuration: the specular reflected light is not collected by the imaging system. Because the imaging system is full field, blocking stray light is essential in obtaining a good signal to noise.

Finally, this single shot camera system is robust against noise and eye motions compared to scanning systems such as OCT.

**Conclusion**

In summary, we demonstrated that our phase imaging device is able to provide cell contrast in transparent cells into the retina *via* a phase to intensity mapping resulting from the oblique trans-scleral illumination.

The full field imaging non-scanning technique combined with transscleral illumination establishes an imaging modality that is several orders of magnitude faster than OCT techniques applied to weak reflecting cells. This is due to the use of the scattered light from the back of the eye which is much more reflective than the transparent cells. The speed of the eye examination is crucial since patients having retinal diseases cannot maintain their eye stable for more than 30 seconds.

Our modality is expected to show promises for clinical use because it is potentially, to our knowledge, the only technique fast enough to image the retina with high contrast and cellular resolution. This imaging method could be complementary to OCT and the combination of both may result in accelerating our understanding of retinal diseases and validating morphological endpoint for therapeutic evaluation, but also the understanding of other neuro degenerative diseases. Recent medical studies have shown that Parkinson and Alzheimer diseases have impact on the neuronal layers of the retina, [40, 41] which may extend the impact of a clinical use of the technology.

## Material and Methods

### Retinal sample preparation
### Retina-Choroid complexes flat-mounting

The human eye sample was obtained from the eye bank of Jules-Gonin eye hospital, in conformity with the Swiss Federal law on the transplantation. It was collected for research purpose at 10 hours *post mortem*, fixed in 4% paraformaldehyde solution (PFA) for 24H, and then stored in 1% PFA at +4°C until analyzes.

Pig eyes were obtained from a local slaughterhouse, a few hours after the death, the eyes were fixed in 4% PFA for 24, and then stored in phosphate-buffered saline (PBS) h at +4°C for one day before mounting.

After washing in PBS, the anterior segment of the eyes and the vitreous were removed. The retina-choroid complexes were detached from the remaining posterior segments, and flat-mounted in a solution of PBS -glycerol (1:1). Human retina-choroid complexes were observed with a confocal microscope (Zeiss LSM710). Human and pig samples were observed with our experimental phase microscope (described in Fig 2d).

### Cryosectioning

One pig eye was kept for cryosectioning. After fixation and microdissection, the posterior segment of the eye was snap- frozen prior to frozen sectioning on a microtome-cryostat. Frontal 10 um-thick sections were prepared and mounted on a glass coverslip. These sections were imaged with a digital holographic microscope (DHM) [43] and then with our phase microscope.

### Immunohistochemistry on neuroretinas

Two three-month-old pigmented rats from the animal facility of Jules-Gonin eye hospital were used. Investigations were performed in accordance of the ARVO statement for the Use of Animals in Ophthalmic Vision Research. The study was approved by the cantonal veterinary office (Authorization VD2928).

Normal retinal morphology was confirmed *in vivo,* under ketamine / xylasine (80mg/kg / 8 mg/kg) anesthesia and after dilation, using a spectral domain OCT system adapted for rat eyes (Bioptigen). Animals were sacrificed by pentobarbital injection. After enucleation and fixation of whole eyes 2 hours in 4% PFA, anterior segments were discarded and neuroretinas were carefully separated from the remaining posterior segment. Post-fixation was carried out for 10 min at -20°C in acetone. After rehydration in PBS supplemented with 0.5% Triton X100 and 10% fetal bovine serum (FBS) overnight at +4°C, retinas were incubated two days at +4°C with polyclonal rabbit anti-NG2 antibody (1:200; AB5320, Merk Millipore), diluted in the same buffer. After washing, neuroretinas were incubated with an Alexa-488-conjugated goat anti rabbit IgG (1:250; Invitrogen), diluted in PBS supplemented with 0.1% Triton X100 and 10% FBS, 2 hours at room temperature. After washing, tissues were stained for 10 min with 4', 6-Diamidino-2-Phenyl-Indole (DAPI, 1:10'000), washed again, and flat-mounted in PBS-glycerol (1:1). Rat flat-mounted neuroretinas were examined with a confocal microscope (Zeiss LSM 710), and then with our phase microscope.

**Study on human subjects**

The single center study aiming at validating the performances of our phase imaging device adheres to the tenets of the Declaration of Helsinki. The protocol involving healthy human subjects was approved by the local Ethics Committee of the Swiss Department of Health on research involving human subjects (CER-VD N°2017-00976). The present study reports the examination of one 49-year-old male subject, presenting with healthy retina.

**Safety consideration**.

A safety analysis of the complete *in vivo* device, including the illumination arm and the optical setup, was carried out. This analysis was approved by an external expert in safety standard for ophthalmic devices. All the doses sent to the eye are safe according to both the European ICNIRP standard [44, 45] and American ANSI standard [46].

**Reconstruction and image processing**.

An acquisition consists in scanning the retina in depth. The shift between each plane is 12 μm. For each z plane, 400 images are recorded (200 from temporal transscleral spot, 200 from the nasal transscleral spot). Then the 80% sharpest images are selected to be averaged. Before averaging, the images are aligned in order to correct for the eye motion thanks to the ImageJ plugin TurboReg that takes into account the translations and the rotations of the image

[42]. Once the images are aligned they are averaged and normalized. Two averaged images, one from left illumination and one from right illumination, are subtracted. A manual shift is performed in order to compute correctly the DPC image. Next, the reconstruction process of the phase is performed [15, 16].

## ACKNOWLEDGEMENTS

The authors thank Dr Michaël Nicolas from the eye bank of Jules-Gonin Eye Hospital for providing a *post-mortem* human eye, Dr Sylvain Roy and Dr Alexandre Matet for fruitful discussions.

This study was supported by a grant from the Technology Transfer Office at EPFL.